\documentclass[superscriptaddress,aps,prl,twocolumn,showpacs,preprintnumbers,amsmath,amssymb,floatfix]{revtex4-1}

\usepackage[normalem]{ulem}

\usepackage{mathrsfs}

\usepackage{graphicx}
\usepackage{dcolumn}
\usepackage{bm}
\usepackage{color}

\def\ba{\begin{eqnarray}}
\def\ea{\end{eqnarray}}
\def\beq{\begin{equation}}
\def\eeq{\end{equation}}

\usepackage[colorlinks,bookmarks=false,citecolor=blue,linkcolor=red,urlcolor=blue]{hyperref}

\begin{document}


\title{Tunable Superfluidity and Quantum Magnetism with Ultracold Polar Molecules}


\author{Alexey V. Gorshkov}
\altaffiliation[]{Authors contributed equally.}
\affiliation{Institute for Quantum Information, California Institute of Technology, Pasadena, California 91125, USA}
\author{Salvatore R. Manmana}
\altaffiliation[]{Authors contributed equally.}
\affiliation{JILA, University of Colorado and NIST, and Department of Physics, CU Boulder, Colorado 80309-0440, USA}
\author{Gang Chen}
\affiliation{JILA, University of Colorado and NIST, and Department of Physics, CU Boulder, Colorado 80309-0440, USA}
\author{Jun Ye}
\affiliation{JILA, University of Colorado and NIST, and Department of Physics, CU Boulder, Colorado 80309-0440, USA}
\author{Eugene Demler}
\affiliation{Physics Department, Harvard University, Cambridge, Massachusetts 02138, USA}
\author{Mikhail D. Lukin}
\affiliation{Physics Department, Harvard University, Cambridge, Massachusetts 02138, USA}
\author{Ana Maria Rey}
\affiliation{JILA, University of Colorado and NIST, and Department of Physics, CU Boulder, Colorado 80309-0440, USA}

\date{\today}

\begin{abstract}

By selecting two 
dressed rotational states of ultracold polar molecules in an optical lattice, we obtain a highly tunable   
generalization of the $t$-$J$ model, which we refer to as the $t$-$J$-$V$-$W$ model.  
In addition to XXZ spin exchange,  
the model features 
density-density interactions 
and 
novel density-spin interactions;  
all interactions are dipolar. 
We  show that full control of all interaction parameters in both magnitude and sign can be achieved independently of each other and of the tunneling. 
As a first step towards demonstrating the potential of the system, we  apply the density matrix renormalization group method (DMRG) to obtain the 1D phase diagram of the simplest experimentally realizable case. Specifically, we show that the tunability and the long-range nature of the interactions in the $t$-$J$-$V$-$W$ model enable enhanced superfluidity.
Finally, we show that Bloch oscillations in a tilted lattice can be used  to probe the phase diagram experimentally. 
\end{abstract} 


\maketitle

Experiments with ultracold 
atoms
have recently extended the range of candidate systems for realizing unconventional states of matter and enabled the 
simulation of models describing condensed matter phenomena \cite{bloch08}.
One major goal of current research at this interface between condensed matter and atomic physics is to emulate the 
Heisenberg  
and $t$-$J$ models, which are believed to underlie 
certain quantum magnetic materials \cite{frustratedQMbook} 
and  high-temperature superconductors \cite{ogata08}, 
respectively. 
However, in the ultracold atom 
realization of these models, 
the small superexchange interaction $J$  \cite{bloch08} 
makes the underlying physics extremely 
challenging to observe.
At the same time, ultracold polar molecules \cite{carr09}, 
such as KRb \cite{ni08,aikawa10}
and LiCs 
\cite{deiglmayr08},
have recently been produced in their electronic and rovibrational 
ground states. 
In this Letter, we show that when such molecules are localized 
in an optical lattice, their rotational degree of freedom can be used to simulate tunable Heisenberg-like models at unit filling of the lattice and, in the presence of doping, a fully tunable generalization of the $t$-$J$ model that we refer to as the anisotropic $t$-$J$-$V$-$W$ model. Dipole-dipole interactions that give rise to this model  are 
orders of magnitude
stronger 
than superexchange interactions $J$ in ultracold atoms and can therefore better compete with other relevant energy and time scales such as, for example, those responsible for decoherence. 
Moreover, we show that the resulting long-range 
interactions are fully controllable with DC electric and continuous-wave microwave fields. As a first step towards demonstrating the potential of the model, we use DMRG \cite{Schollwock_RMP} to obtain the 1D phase diagram for the simplest experimentally relevant case and show that, at low fillings,  
the 
superfluid phase is enhanced relative to the one in the conventional $t$-$J$ model \cite{moreno11}.
We propose to probe the phase diagram using center-of-mass Bloch oscillations. 
Given that KRb has already been loaded into a 3D lattice, our proposal is 
applicable to current experiments. 

\textit{The Hamiltonian and its features.}---We consider diatomic polar molecules in their electronic and vibrational ground state  partially polarized by a DC electric field along $\mathbf{\hat z}$, confined to the $x$-$y$ plane  \cite{demiranda10},  and loaded in that plane into the lowest band of a square optical lattice. 
As described below, microwave fields are used to isolate in each molecule two dressed rotational states $|m_0\rangle$ and $|m_1\rangle$ and to obtain the $t$-$J$-$V$-$W$ Hamiltonian
\begin{eqnarray}
&&H =  - t \sum_{\langle i,j\rangle m}  \left[c^\dagger_{i m} c_{j m} + \textrm{h.c.}\right]+ \sum_{i \neq j} |\mathbf{R}_i-\mathbf{R}_j|^{-3} 
\nonumber\\
&&\times \Bigg[  \frac{J_\perp}{2} S^+_i S^-_j  + \frac{J_z}{2} S^z_{i} S^z_{j} + \frac{V}{2} n_i n_j + W n_i S^z_{j}  \Bigg].  \label{eq:tjvw} 
\end{eqnarray} 
The two terms  describe tunneling and dipole-dipole interactions, respectively;  $\langle \rangle$ denotes nearest-neighbor bonds.
Specifically, $c^\dagger_{j m}$ creates a fermionic \cite{ni08} 
or bosonic  \cite{deiglmayr08,aikawa10} 
molecule on site $j$ (position $\mathbf{R}_j$) in dressed rotor state $m \in \{m_0,m_1\}$; in this Letter, we focus on fermions.
Large reaction rates  \cite{demiranda10} 
between two molecules on the same site 
enforce the hardcore constraint.
We define $n_{j m}= c^\dagger_{j m} c_{j m}$, $n_j = \sum_m n_{j m}$,  
$S_j^+ =  c^\dagger_{j m_0} c_{j m_1}$, $S_j^z = (n_{jm_0}-n_{jm_1})/2$ and use units in which $\hbar = 1$. 
The $J_z$, $V$, and $W$ terms can be understood by thinking of $|m_0\rangle$ and $|m_1\rangle$ as classical permanent dipoles oriented along $\mathbf{\hat z}$, while the $|m_0\rangle$-$|m_1\rangle$ transition dipole moment gives rise to the $J_\perp$ term.
We tune these dipole moments by constructing dressed states 
$|m_0\rangle$ and $|m_1\rangle$ out of bare rotor states using microwave fields.
This, in turn, allows for the full controllability of $J_z$, $J_\perp$, $V$, and $W$, which is one of the main results of the present Letter. 
Lattice Hamiltonians based on more than one molecular rotational state 
have been considered before in Refs.\ 
 \cite{barnett06, micheli06,*brennen07,*buchler07b, watanabe09, wall09, yu09b, krems09,wall10,schachenmayer10,perezrios10,trefzger10,herrera10,kestner11}.   
An important difference of Eq.\ (\ref{eq:tjvw}) 
from the Hamiltonian studied in Ref.\ \cite{wall10}, which is most closely related to our work, is the presence of the $J_\perp$ term \cite{barnett06}.

Eq.~(\ref{eq:tjvw}) possesses several 
aspects that can lead to unconventional many-body phases.
Most notably, all the 
interactions are long-range;
in particular, 
repulsive long-range density-density interactions are expected to stabilize 
superfluid correlations 
\cite{troyer93}.
Furthermore, the novel 
$W$ 
term 
can 
break SU(2) symmetry of the system even when $J_z = J_\perp$.
It is also crucial that $J_z$ and $J_\perp$  can be tuned in both sign and magnitude (up to $\sim 100$ kHz in LiCs) independently from each other and from $t$. 
This contrasts with the cold atom realization of the $t$-$J$ model, where $J \ll t$.  
Finally, we note that, in the limit of unit filling ($n=1$), only the terms with $J_z$ and $J_\perp$ survive, realizing an XXZ-model with long-range 
dipolar interactions, and, for $J_z = J_\perp$, the Heisenberg model.

\begin{figure}[t]
\begin{center}
\includegraphics[width = 0.98 \columnwidth]{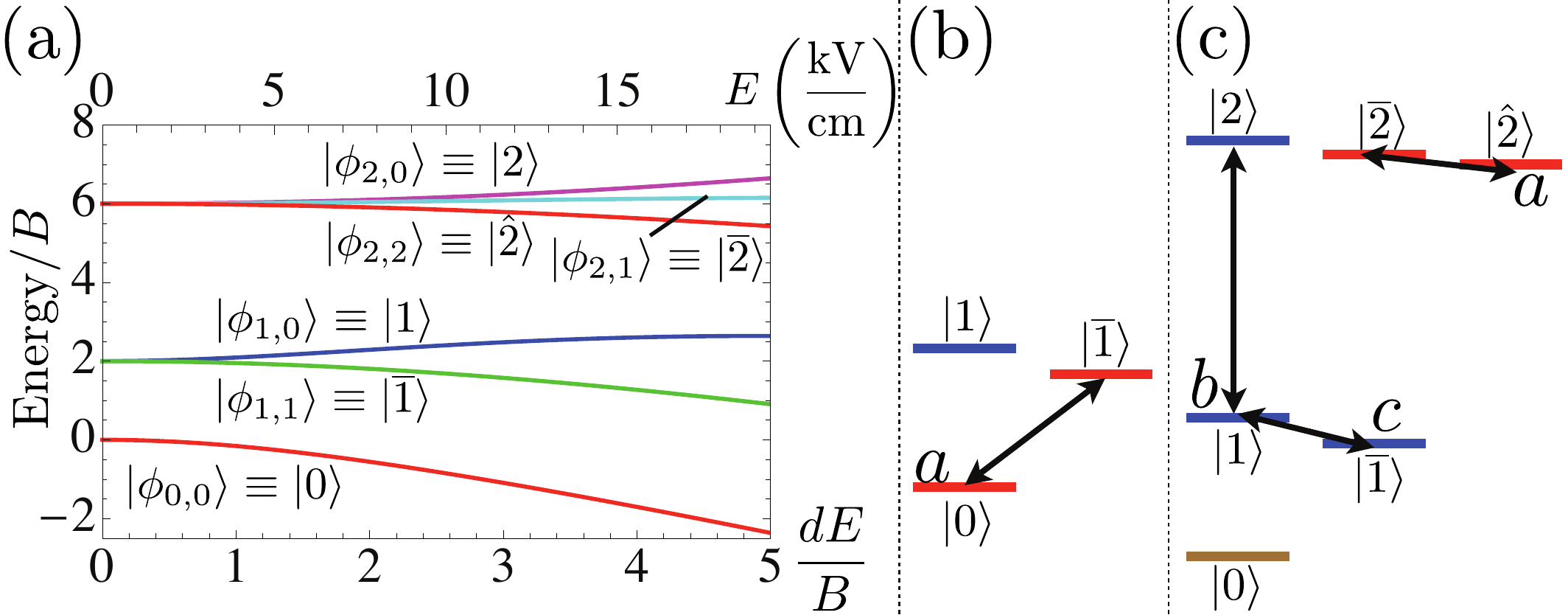}
\caption{(color online). (a) Eigenenergies of $H_0 = B \mathbf{N}^2 - d_0 E$ 
as a function of $E$ (top axis uses $d$ and $B$ of KRb). Only $M \geq 0$ states are labeled. (b,c) Examples of level configurations $\{|m_0\rangle,|m_1\rangle\}$: (b)  $\{ \sqrt{a} |0\rangle + \sqrt{1-a} |\overline{1}\rangle,  |1\rangle\}$, (c) $\{\sqrt{a} |\hat{2}\rangle + \sqrt{1-a} |\overline{2}\rangle,\sqrt{b} |1\rangle + \sqrt{c} |\overline{1}\rangle + \sqrt{1-b-c} |2\rangle\}$.  Red (blue) levels make up the dressed rotor state $|m_0\rangle$ ($|m_1\rangle$). \label{fig:levels}}
\end{center}
\end{figure}

\textit{Tuning  $J_z$, $J_\perp$, $V$, and $W$.}---
We now show how to use DC electric and microwave fields \cite{micheli06,*brennen07,*buchler07b,buchler07,micheli07,gorshkov08c,yu09b,krems09,lin10b,cooper09,wall09,wall10,schachenmayer10,kestner11} 
 to control $J_z$, $J_\perp$, $V$, and $W$, 
leaving a detailed derivation of Eq.~(\ref{eq:tjvw}) 
to Ref.~[\onlinecite{gorshkov11b}]. 
A polar molecule in a DC electric field is described by the rigid-rotor Hamiltonian $H_0 = B \mathbf{N}^2 - d_0 E$, with  the rotational constant $B$, the angular momentum operator $\mathbf{N}$, and the dipole moment operator $\mathbf{d}$. We set $d_p = \hat{\mathbf{e}}_p \cdot \mathbf{d}$, where 
$\mathbf{e}_0 = \mathbf{z}$ and $\mathbf{e}_{\pm} = \mp (\mathbf{x} \pm i \mathbf{y})/\sqrt{2}$ \cite{micheli07}. 
At $E = 0$, $H_0$ has eigenstates $|N,M\rangle$ obeying $\mathbf{N}^2 |N,M\rangle = N (N+1) |N,M\rangle$ and $N_z |N,M\rangle = M |N,M\rangle$. 
As $E$ is increased, eigenstates with the same $M$ mix, forming superpositions of $|N,M\rangle$ with different $N$. 
We denote the state at finite $E$ adiabatically connected to $|N,M\rangle$ as $|\phi_{N,M}\rangle$ and define, for notational convenience, $|N\rangle \equiv |\phi_{N,0}\rangle$, $|\overline{N}\rangle \equiv |\phi_{N,1}\rangle$, and  $|\hat{N}\rangle \equiv |\phi_{N,2}\rangle$ [see Fig.\ \ref{fig:levels}(a)]. 
For appropriate choices of 
levels,
energy conservation ensures that dipole-dipole interactions between two molecules preserve their total $N_z$, resulting in the many-body interaction Hamiltonian  
$H_\textrm{dd} = \frac{1}{2} \sum_{i \neq j} |\mathbf{R}_i-\mathbf{R}_j|^{-3} 
[d_0^{(i)} d_0^{(j)} + \frac{1}{2} (d_+^{(i)} d_-^{(j)} + d_-^{(i)} d_+^{(j)})]$. 
We now focus on the two configurations of levels and microwave fields shown in Figs.\ \ref{fig:levels}(b,c), which allow, respectively, for the realization of the simple case $J_z = V = W = 0$ and for the full tunability of $J_z$, $J_\perp$, $V$, and $W$.
  
In the first configuration, states $|m_0\rangle = \sqrt{a} |0\rangle + \sqrt{1-a} |\overline{1}\rangle$ and $|m_1\rangle = |1\rangle$ [Fig.\ \ref{fig:levels}(b)] are chosen, where the coefficient $a$ in the dressed state $|m_0\rangle$ is controlled by the ratio between the Rabi frequency and the detuning of a $\sigma_+$-polarized microwave field acting on the $|0\rangle$-$|\overline{1}\rangle$ transition \cite{gorshkov08c}. In ${}^{40}$K${}^{87}$Rb, microwave intensity of a few W/cm$^2$ is sufficient to 
address all hyperfine levels equally.
Projecting on states $|0\rangle$, $|1\rangle$, and $|\overline{1}\rangle$ and keeping only energy conserving terms, we obtain 
$d_0^{(i)} d_0^{(j)} =  \mu_{01}^2 |01\rangle \langle 10|_{ij} + \textrm{h.c.} + \otimes_{k} \sum_{s}  \mu_s |s\rangle \langle s|_k$
and $d_+^{(i)} d_-^{(j)} + d_-^{(i)} d_+^{(j)} =   - \mu_{0 \overline{1}}^2 |0 \overline{1}\rangle \langle \overline{1}0|_{ij}  - \mu_{1 \overline{1}}^2 |1 \overline{1}\rangle \langle \overline{1}1|_{ij} + \textrm{h.c.}$,
where $k \in \{i,j\}$, $s \in \{0,1,\overline{1}\}$, $\mu_{s} = \langle s|d_0|s\rangle$, and $\mu_{ss'} = \langle s|d_p|s'\rangle$ for the appropriate $p$.
The minus sign in front of $\mu_{0 \overline{1}}^2$ and $\mu_{1 \overline{1}}^2$ is crucial to the tunability of $V$, $W$, $J_z$, and $J_\perp$ and appears because 
two dipoles rotating in the $x$-$y$ plane interact on average attractively \cite{gorshkov08c}. 
Projecting on $|m_0\rangle$ and $|m_1\rangle$ and again keeping only energy conserving terms, we obtain 
\ba
&& d_0^{(i)} d_0^{(j)} + \tfrac{1}{2} (d_+^{(i)} d_-^{(j)} + d_-^{(i)} d_+^{(j)}) = \sum_p B_p |m_p m_p\rangle \langle m_p m_p|_{ij}   \nonumber \\
&&+ \!\!  \sum_{p,q} \! A_p A_q |m_p m_q\rangle \langle m_p m_q|_{ij} \!+\! \frac{J_\perp}{2} (|m_0 m_1\rangle \langle m_1 m_0|_{ij} \!+\! \textrm{h.c.}\!), \label{eq:dd} \nonumber
\ea
where $p,q \in \{0,1\}$ and 
$A_0 = a \mu_0 + (1-a) \mu_{\overline{1}}$, $A_1 = \mu_1$, $B_0 = -\mu_{0 \overline{1}}^2 a (1-a)$, $B_1 = 0$, $J_\perp = 2 \mu_{01}^2 a - \mu_{1\overline{1}}^2 (1-a)$.  $A_p$ can be thought of as the effective dipole moment of $|m_p\rangle$, while $B_p$ is the contribution to $H_{dd}$ from transition dipole moments between rotor states making up $|m_p\rangle$. From the comparison of 
this equation and Eq.\ (\ref{eq:tjvw}), we can read off $V = [(A_0+A_1)^2 + B_0 + B_1]/4$, $W = (A_0^2 +B_0 - A_1^2  - B_1)/2$, $J_z = (A_0-A_1)^2 + B_0+B_1$. A simple case $J_z = W = V = 0$ and $J_\perp > 0$ studied below can be approximately implemented using $a = 1$, which does not require a microwave field. A small $E$ field ($d E/B > 0.1$ in ${}^{40}$K${}^{87}$Rb) 
is needed to prevent dipole-dipole 
and 
hyperfine interactions from populating $|\phi_{1,\pm 1}\rangle$. At $d E/B = 0.1$, 
 $V$, $W$, and $J_z$ are two orders of magnitude smaller than $J_\perp$ and can be neglected. 

The second configuration we consider is $\{|m_0\rangle,|m_1\rangle\} = \{\sqrt{a} |\hat{2}\rangle + \sqrt{1-a} |\overline{2}\rangle,\sqrt{b} |1\rangle + \sqrt{c} |\overline{1}\rangle + \sqrt{1-b-c} |2\rangle\}$ [Fig.\ \ref{fig:levels}(c)]. 
The three microwave fields shown in the figure, which allow to control the coefficients $a$, $b$, and $c$, together with the DC electric field constitute four knobs that allow for the full control over the four coefficients $V$, $W$, $J_z$, and $J_\perp$  
\footnote{In this configuration, $J_\perp = 2 \mu_{{\overline 2}{\overline 1}}^2 c (1-a) - \mu_{{\overline 2}1}^2 (1-a) b - \mu_{{\overline 2}2}^2 (1-a) (1-b-c) - \mu_{{\hat 2} {\overline 1}}^2 a c$, $A_0 = \mu_{\hat 2} a + \mu_{\overline 2} (1-a)$, $A_1 = \mu_1 b + \mu_{\overline 1} c + \mu_2 (1-b-c)$, $B_0 = -\mu_{{\hat 2}{\overline 2}}^2 a (1-a)$, and $B_1 = 2 \mu_{12}^2 b (1-b-c) - \mu_{1{\overline 1}}^2 b c - \mu_{2{\overline 1}}^2 c (1-b-c)$.}. 
We find that in a small sphere in the 4-dimensional $(dE/B,a, b,c)$ space  around the point $(d E/B,a,b,c) = (2.97,0.059, 0.56,0.38)$, where $J_z = J_\perp = V = W = 0$, one can achieve any value of $V$, $W$, $J_z$, and $J_\perp$ up to an overall positive prefactor. Similarly, in the special case where a single microwave couples $|\overline{1}\rangle$ and $|1\rangle$, we have a two-dimensional $(d E/B,b)$ subspace (with $a = 0$, $c = 1 - b$), in which $J_z$ and $J_\perp$ can be fully controlled -- for simulations of the XXZ model at unit filling -- around $(d E/B,b) = (4.36,0.56)$, where they both vanish.
While these examples prove full controllability in their respective cases, for any desired relationship between $V$, $W$, $J_z$, and $J_\perp$, there is likely a different level configuration that gives stronger interactions and uses weaker $E$, lower microwave intensity, and/or more convenient microwave frequencies.



\begin{figure}
\includegraphics[width=0.7\columnwidth]{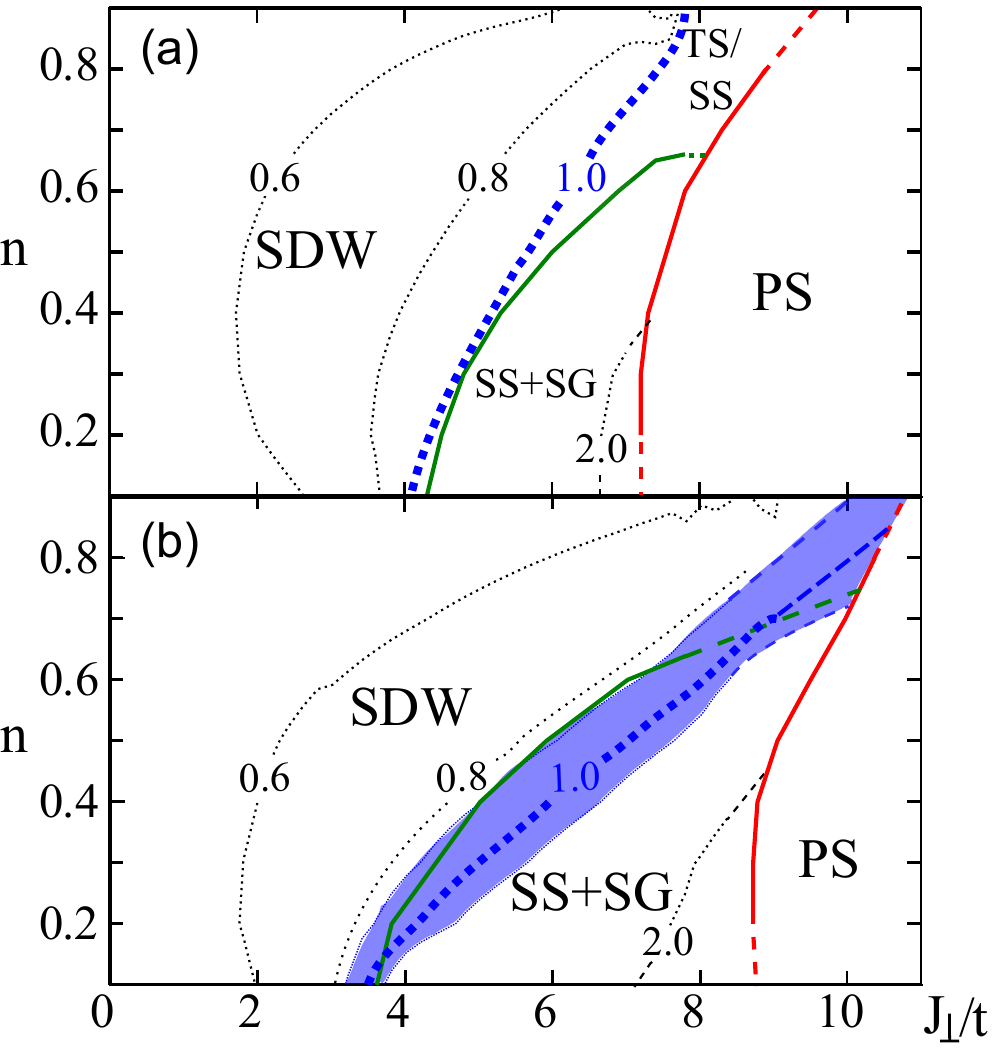}
\caption{(Color online).\  Phase diagrams of the $t$-$J_\perp$ chain  with (a) nearest-neighbor and (b) dipolar 
$J_\perp$. 
We identify 
a metallic (repulsive Luttinger liquid) phase 
with dominant spin-density-wave correlations (SDW), a gapless superfluid with dominant triplet and singlet superfluid correlations (TS/SS), 
a singlet superfluid with a spin gap (SS+SG), 
and phase separation (PS). The spin gap is $ \lesssim 0.35 t$ in (a) and $ \lesssim 0.7t$ in (b).
Solid lines indicate phase transitions (green: SG closes; red: inverse compressibility becomes zero);  
dashed lines are extrapolations. The numbers show the value of $K_\rho$ on the dotted lines. 
The line $K_\rho = 2$ is a crossover line within the SS+SG phase \cite{giamarchi04}.
The shaded region in (b) displays $K_\rho = 1 \pm 0.15$ as an estimate of the numerical accuracy.
}
\label{fig:phasedias}
\end{figure}

\textit{Phase diagrams of the nearest-neighbor and dipolar $t$-$J_{\perp}$ chains.}---The full tunability of the $t$-$J$-$V$-$W$ model provides access to a great variety of models with potentially exotic physics. As the simplest example of this physics, 
we present in Fig.~\ref{fig:phasedias} the 1D phase diagram in the limit 
$V = W = J_z  = 0$, which  
 is one of the simplest experimentally achievable cases (see above). 
Before analyzing 
dipolar interactions,  we present in Fig.~\ref{fig:phasedias}(a) the phase diagram of the \textit{nearest-neighbor} $t$-$J_\perp$ chain as obtained using 
DMRG 
and following the analysis of Ref.~[\onlinecite{moreno11}].  
The diagram is 
qualitatively similar to that of the standard $t$-$J$ chain \cite{moreno11}: 
   At fillings $n < 0.65$, we identify a repulsive Luttinger liquid (Luttinger parameter $K_\rho < 1$) with dominant spin-density-wave (SDW) correlations,  an attractive Luttinger liquid ($K_\rho > 1$) with dominant singlet and triplet superfluid correlations, a singlet superfluid with a spin gap, and phase separation.  
At larger fillings, the spin gap is always zero, but the other phases remain.   

In Fig.~\ref{fig:phasedias}(b), we present the phase diagram of this system in the presence of 
dipolar interactions. 
At low fillings, SDW, the gapped singlet superfluid, and phase separation are obtained. 
This suggests that experiments with ultracold 
molecules  
can be 
a useful tool for exploring the phase diagram of the standard $t$-$J$ model for {\it arbitrary} values of $J$ in contrast to $J \ll t$ in 
ultracold atom realizations.  
At the same time, crucially, both diagrams in Fig.\ \ref{fig:phasedias} feature a significant enhancement of  the 
superfluid region compared to the original $t$-$J$ model because the absence of attractive density-density interactions 
suppresses phase separation. Furthermore, the maximum value of the spin gap in the dipolar $t$-$J_{\perp}$ chain is twice that in the nearest-neighbor $t$-$J_{\perp}$ chain, which is, in turn, twice that in the original $t$-$J$ model. The larger spin gap should facilitate its experimental observation.
Note that, as a conservative estimate of the numerical errors in computing $K_\rho$ in the presence of long-range interactions with up to 100 sites, we estimate the true line $K_\rho = 1$ to lie  in the shaded region $0.85 \leq K_\rho \leq 1.15$.
Thus, 
the gapless superfluid 
 cannot be identified in Fig.~\ref{fig:phasedias}(b) within our numerical precision. 
Furthermore, while 
the line $K_\rho = 1$ and the line where the spin gap closes may coincide, our analysis of the correlation functions cannot rule out the existence of an exotic intermediate phase with a spin gap, $K_\rho < 1$, and dominant 
superfluid correlations. 

\begin{figure}[b]
\includegraphics[width=0.7\columnwidth]{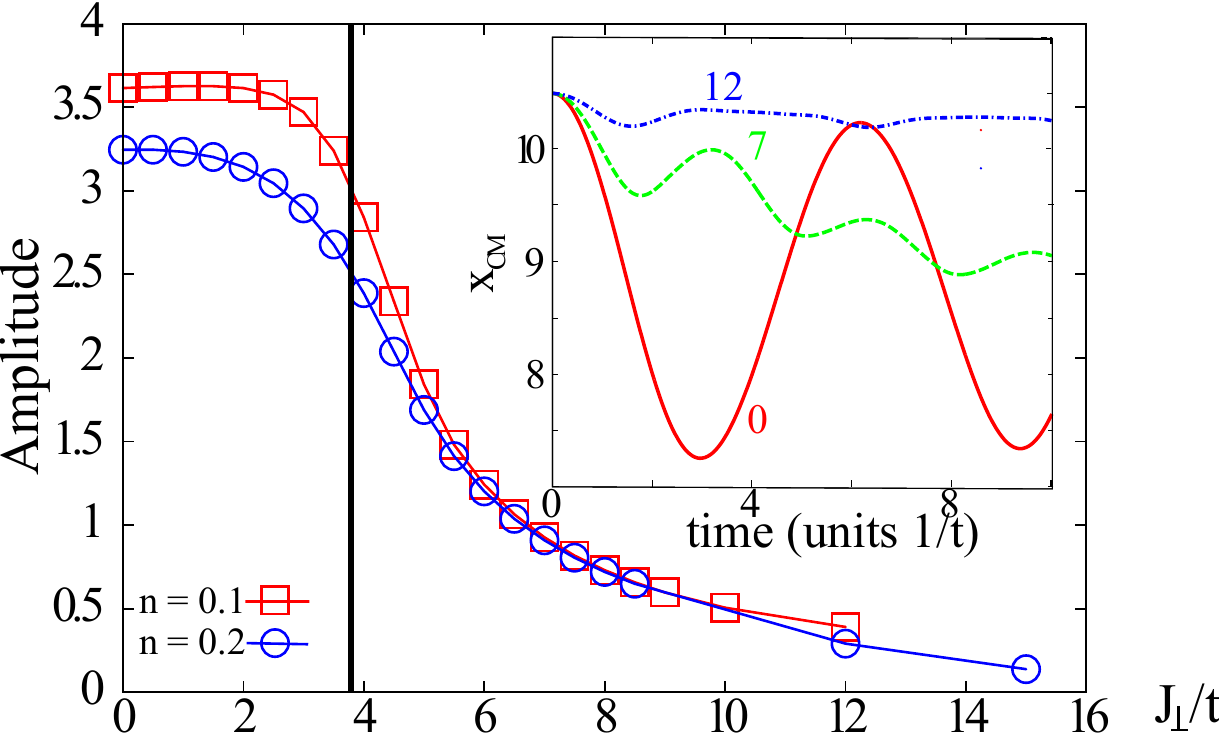} 
\caption{(color online).\ 
Bloch oscillations in a dipolar $t$-$J_\perp$ chain of 
20 sites with open boundary conditions at fillings $n=0.1$ and $0.2$, and tilting field $E_{\rm tilt}/t = 1$ 
per site. 
The main plot shows the difference between the center-of-mass position $x_{\rm CM}$ at time zero and at the first minimum.
The vertical line indicates the approximate value of $J_\perp/t$, at which the spin gap closes 
for $n =0.2$. The inset shows the time evolution of $x_{\rm CM}$ at $n=0.2$ at three indicated values of $J_\perp/t$. 
\label{fig:Blochoscillations}}
\end{figure}

\textit{Preparation and detection.}--- Ground-states at specific points in the phase diagram can be prepared, e.g., by applying an additional microwave field coupling $|m_0\rangle$ and $|m_1\rangle$ and performing an adiabatic passage from an easily accessible state to the desired ground state by tuning the Rabi frequency and the detuning of the microwave field \cite{schachenmayer10}. 
Direct probing of molecules \cite{wang10} and conversion of molecules back to atoms \cite{ni08} can in principle both be used for detection via
noise-correlations in the time-of-flight 
\cite{bloch08} 
or via in-situ single-site 
imaging \cite{bakr10,sherson10}. 

As a specific example of a detection technique available in current experiments, we propose to use Bloch oscillations \cite{ben-dahan96,wilkinson96}. 
In Fig.~\ref{fig:Blochoscillations}, we present our results as obtained via the Krylov-space variant of the adaptive t-DMRG \cite{Schollwock_MPS,Manmana:2005p63} for 
20 sites when adding a linear field along the chain  
at filling 
$n=0.1$ and $n=0.2$ for different values of $J_{\perp}/t$. 
In the singlet superfluid, 
the amplitude of the oscillations drops and the frequency of the oscillations doubles  (see $J_\perp/t = 7$ in the inset) relative to the gapless phase due to the presence of bound pairs.  
Bloch oscillations should be observable in direct absorption imaging \cite{wang10}. 
While neither the frequency nor the amplitude of the oscillations show any sharp features at this small system size, the fit of experimental data to numerical results should allow for the location of the phase transition (vertical line in Fig.~\ref{fig:Blochoscillations}) even for small system sizes.
As a complementary method for identifying the transition, we propose spectroscopic measurement of the  spin gap \cite{buchler04,*guo10}.
\textit{Outlook.}---We have presented a toolbox for simulating a highly tunable anisotropic $t$-$J$-$V$-$W$ model with polar molecules. The advantages of this molecular toolbox over its atomic counterpart are higher energy scales and independent tunability of interactions and tunneling. This toolbox should enable the simulation of condensed matter phenomena, as well as the stabilization and controllable preparation of unconventional phases, such as $d$-wave superfluids  \cite{ogata08}. 
The phase diagram of the experimentally simple case of a $t$-$J_\perp$ chain shows an enhanced superfluid 
region, which we propose to probe via Bloch oscillations. We expect that the 2D $t$-$J$-$V$-$W$ model can similarly be tuned into exhibiting enhanced superfluidity \cite{dagotto92}.

The present Letter  also opens other exciting research avenues. In particular, natural extensions of the model include \cite{gorshkov11b}: spatially anisotropic interactions produced by a tilt in the DC electric field, spin-dependent tunneling obtained by adjusting lattice beams, $S > 1/2$ models realized by choosing more than two dressed states, and systems with an orbital degree of freedom encoded in the nuclear spin. Moreover,  by considering molecular Wigner crystals \cite{buchler07,rabl07}, where the intermolecular distances are smaller than in an optical lattice, one can further increase the interaction strength. Furthermore, by analogy with Ref.\ \cite{schachenmayer10}, we expect our ideas to be extendable to Rydberg atoms.  Finally, 
one can envision applications of the present system to quantum computation (especially if one uses nuclear spin to store information), precision measurements, and controlled quantum chemistry 
\cite{carr09}.

We thank A.\ Muramatsu, M.\ Troyer, A.\ Moreno, D.\ Jin, G.\ Refael, J.\ Aldegunde, 
P.\ Julienne, M.\ Babadi,  I.\ Bloch, 
P.\ Rabl, G.\ Qu\'em\'ener, A.\ Potter, B.\ Wunsch, P.\ Zoller, S.\ Pielawa, E.\ Berg, A.\ Daley, F.\ Mila, R.\ Noack, and H.\ Weimer 
 for discussions. This work was supported by the NSF, NIST, 
 the Lee A.\ DuBridge Foundation,
  the ARO with funding from DARPA-OLE, CUA, and AFOSR MURI.
We acknowledge CPU time at ARSC. 




\bibliography{refs}

\end{document}